\documentclass[letterpaper, 10 pt, conference]{ieeeconf}  

\IEEEoverridecommandlockouts                              

\usepackage{amsmath} 
\usepackage{amssymb}  
\usepackage{graphicx,adjustbox,etoolbox,hyperref}

\graphicspath{{./figures/}}

\title{\LARGE \bf
MedNeRF: Medical Neural Radiance Fields for\\ Reconstructing 3D-aware CT-Projections from a Single X-ray
}

\author{Abril Corona-Figueroa$^{1}$, Jonathan Frawley$^{1}$, Sam Bond-Taylor$^{1}$,\\ Sarath Bethapudi$^{2}$, Hubert P. H. Shum$^{1}$, Chris G. Willcocks$^{1}$
\thanks{$^{1}$Corona-Figueroa, Frawley, Bond-Taylor, Shum and Willcocks are with the Computer Science Department, Durham University,
        Durham, DH1 3LE, UK
        {\tt\small abril.corona-figueroa@durham.ac.uk}}%
\thanks{$^{2}$Bethapudi is with the County Durham and Darlington NHS Foundation Trust,
        Durham, DL3 6HX, UK}%
}

\makeatletter
\patchcmd{\@makecaption}
  {\scshape}
  {}
  {}
  {}
\makeatletter
\patchcmd{\@makecaption}
  {\\}
  {.\ }
  {}
  {}
\makeatother

\begin{document}

\maketitle
\thispagestyle{empty}
\pagestyle{empty}

\begin{abstract}

Computed tomography (CT) is an effective medical imaging modality, widely used in the field of clinical medicine for the diagnosis of various pathologies. Advances in Multidetector CT imaging technology have enabled additional functionalities, including generation of thin slice multiplanar cross-sectional body imaging and 3D reconstructions. However, this involves patients being exposed to a considerable dose of ionising radiation. Excessive ionising radiation can lead to deterministic and harmful effects on the body. This paper proposes a Deep Learning model that learns to reconstruct CT projections from a few or even a single-view X-ray. This is based on a novel architecture that builds from neural radiance fields, which learns a continuous representation of CT scans by disentangling the shape and volumetric depth of surface and internal anatomical structures from 2D images. Our model is trained on chest and knee datasets, and we demonstrate qualitative and quantitative high-fidelity renderings and compare our approach to other recent radiance field-based methods. Our code and link to our datasets are available at \url{https://github.com/abrilcf/mednerf}

\indent \textit{Clinical relevance}— Our model is able to infer the anatomical 3D structure from a few or a single-view X-ray, showing future potential for reduced ionising radiation exposure during the imaging process.
\end{abstract}

\section{INTRODUCTION}

3D medical imaging often involves joining multiple 2D slices from CT or Magnetic Resonance Imaging (MRI), and part of their workflow consists of specifying values for the position of the patient, the imaging source, and the detector. The quality and accuracy of a CT 3D representation require hundreds of X-ray projections with a thin slice thickness \cite{suetens_2009}. Moreover, this process exposes patients to more ionising radiation than typical X-rays and requires the patient to remain immobile for up to more than 1 hour, depending on the type of test \cite{pechin_2012}. Continuous 3D representations would give radiologists optics of every point in the internal anatomy captured. While such representations are useful, there are practical challenges in CT due to the increased radiation exposure, angle-dependent structures, and time consumption \cite{COFFEY_2010}. 

Earlier approaches in medical image reconstruction used analytic and iterative methods \cite{Huynh_2016,xie_2020} on given input data. However, they often encounter mismatches between the mathematical model and physical properties of the imaging system. Instead, several recent approaches leverage deep learning \cite{Wang_2018} for sparse view reconstruction \cite{Li_2019,autoint_2021,coil_2021}, 3D CT reconstruction from 2D images \cite{Ying_2019}, and anomaly detection \cite{nguyen_2021}. These deep learning approaches solved the mismatches between the mathematical model and imaging system and reported improved reconstructions by fine-tuning state-of-the-art architectures. However, they require a large amount of training data, which may be difficult to meet in the medical domain where acquiring expert annotations is both cost and time prohibitive.

The Neural Radiance Fields (NeRF) \cite{nerf_2020} model is a \textit{recent} reformulation for estimating a 3D volumetric representation from images. Such representations encode the radiance field and density of the scene in the parameters of a neural network. The neural network learns to synthesize new views via volume rendering from point samples along cast rays. However, these representations are often captured in controlled settings \cite{2020nerfw}. First, the scene is taken by a set of fixed cameras within a short time frame. Second, all content in the scene is static and real images often need masking. These constraints prohibit the direct application of NeRF to the medical domain, where the imaging system greatly differs from conventional cameras, and the images are captured over a long time frame hampering the patient's stillness. Moreover, the overlapping of anatomical structures in medical images hinders the definition of edges which cannot be easily solved with masking. These aspects explain why the NeRF approach especially shows successes for ``natural images''.

To address these challenges, we propose MedNeRF, a model that adapts Generative Radiance Fields (GRAF) \cite{Schwarz2020NEURIPS} in the medical domain to render CT projections given a few or even a single-view X-ray. Our approach not only synthesizes realistic images, but also capture the data manifold and provides a continuous representation of how the attenuation and volumetric depth of anatomical structures vary with the viewpoint without 3D supervision. This is achieved via a new discriminator architecture that provides a stronger and more comprehensive signal to GRAF when dealing with CT scans.

Closest to our goal are \cite{autoint_2021,coil_2021}, which both train a coordinate-based network in sinograms of low-dose CT of phantom objects and apply it to the sparse-view tomography reconstruction problem. In contrast to \cite{autoint_2021}, we learn multiple representations in a single model by randomly feeding data of different medical instances instead of separately optimizing for each collection of images. For testing \cite{coil_2021} reconstruction ability, they integrate it into reconstruction methods and use at least 60 views. Different from their methods, we do not rely on additional reconstruction algorithms, and we only require multiple views during training.

We render CT projections of our two datasets of digitally reconstructed radiographs (DRR) from chest and knee. We qualitative and quantitative demonstrate high-fidelity renderings and compare our approach to other recent radiance field-based methods. Furthermore, we render CT projections of a medical instance given a single-view X-ray and show the effectiveness of our model to cover surface and internal structures.

\section{METHODS}
\subsection{Dataset Preparation}

To train our models, we generate DRRs instead of collecting paired X-rays and corresponding CT reconstructions, which would expose patients to more radiation. Furthermore, DRR generation removes patient data and enables control in capture ranges and resolutions. We generated DRRs by using 20 CT chest scans from \cite{Tsai_2021,TCIA_2013} and five CT knee scans from \cite{Harris_2016,Ali_2016}. These scans cover a diverse group of patients at different contrast types showing both normal and abnormal anatomy. The radiation source and imaging panel are assumed to rotate around the vertical-axis, generating a DRR of \(128 \times 128 \) resolution at every five degrees, resulting in 72 DRRs for each object. During training we use the whole set of 72 DRRs (a fifth of all views within a full 360-degree vertical rotation) per patient and let the model render the rest. Our work did not involve experimental procedures on human subjects or animals and thus did not require Institutional Review Board approval.

\subsection{GRAF Overview}

GRAF \cite{Schwarz2020NEURIPS} is a model that builds from NeRF and defines it within an Generative Adversarial Network (GAN). It consists of a generator \(G_\theta\) that predicts an image patch \(\boldsymbol{P}_\textup{pred}\) and a discriminator \(D_\phi\) that compares the predicted patch to a patch \(\boldsymbol{P}_\textup{real}\) extracted from a real image.
GRAF has shown an effective capacity to disentangle 3D shape and viewpoint of objects from 2D images alone, in contrast to the original NeRF \cite{nerf_2020} and similar approaches such as \cite{yu2020pixelnerf}. Therefore, we aim to translate GRAF's methods to our task, and in subsection \ref{mednerf} we describe our new discriminator architecture, which allows us to disentangle 3D properties from DRRs.

\begin{figure}[t]
\centering
\setlength\fboxsep{0pt}
\setlength\fboxrule{0.25pt}
\includegraphics[width=8.5cm]{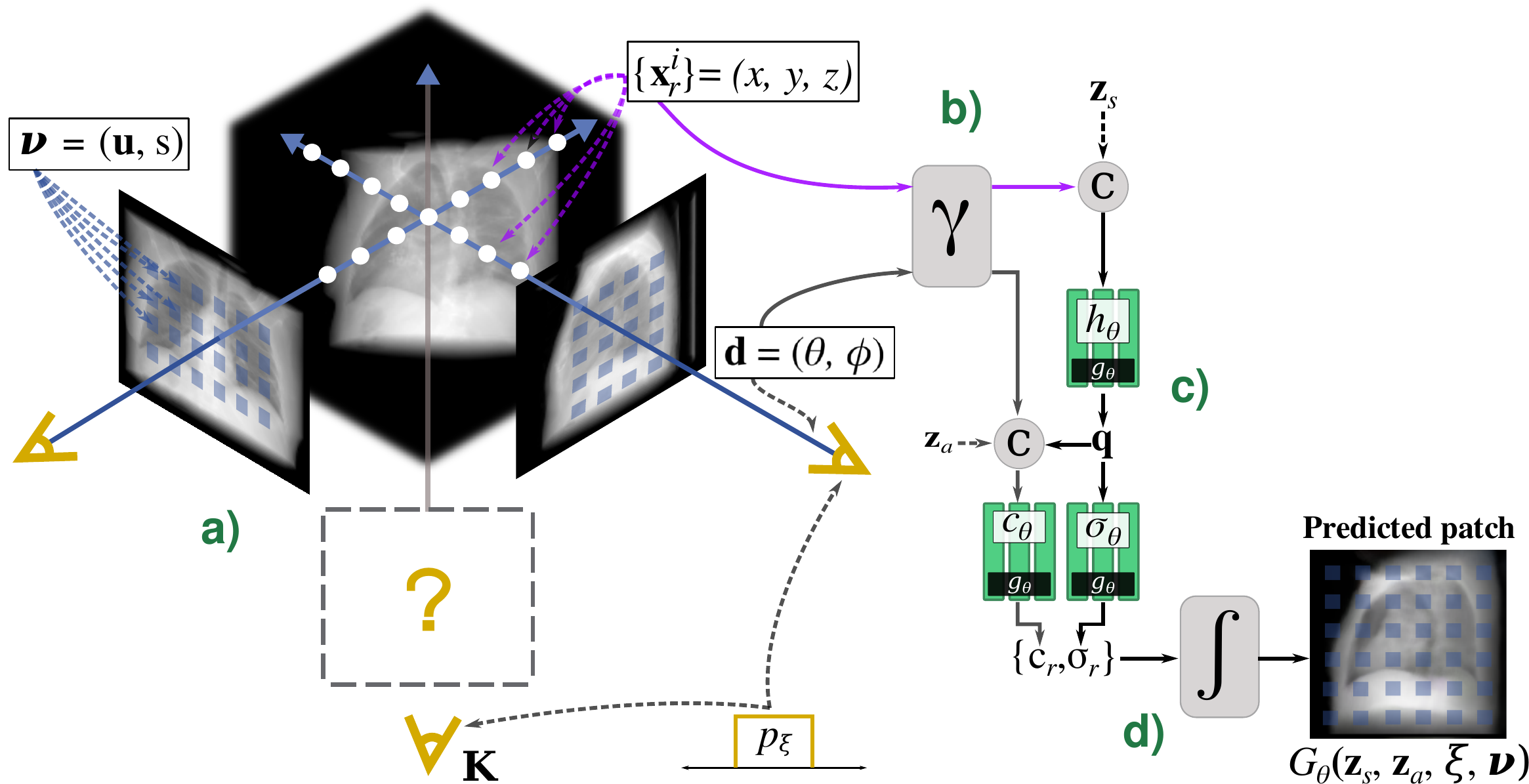}
\caption{An overview of GRAF's generator.} 
\label{generator}
\end{figure}

We consider the experimental setting to obtain the radiation attenuation response instead of the color used in natural images. To obtain the attenuation response at a pixel location for an arbitrary projection \(\boldsymbol{K}\) with pose \(\boldsymbol{\xi}\), first, we consider a pattern \(\boldsymbol{\nu}=(\boldsymbol{u},s)\) to sample \(R\) X-ray beams within a \(K\times K\) image-patch \(\boldsymbol{P}\). Then, we sample \(N\) 3D points \({\boldsymbol{x}^i_r}\) along the X-ray beam \(\boldsymbol{r}\) originating from the pixel location and ordered between the near and far planes of the projection (Fig. \ref{generator}a).

The object representation is encoded in a multi-layer perceptron (MLP) that takes as input a 3D position \(\boldsymbol{x}=(x,y,z)\) and a viewing direction \(\boldsymbol{d}=(\theta,\phi)\), and produces as output a density scalar \(\sigma\) and a pixel value \(c\).  To learn high-frequency features, the input is mapped into a \(2L\)-dimensional representation (Fig. \ref{generator}b):
\begin{equation}
        \gamma(p)= ...,\cos(2^j\pi p),\sin(2^j\pi p),...
\end{equation}
where \(p\) represents the 3D position or viewing direction, for \(j = 0,...,m-1\).

For modeling the shape and appearance of anatomical structures, let \(\boldsymbol{z}_s\!\sim\! p_s\) and \(\boldsymbol{z}_a\!\sim\!p_a\) be the latent codes sampled from a standard Gaussian distribution, respectively (Fig. \ref{generator}c). To obtain the density prediction \(\sigma\), the shape encoding \(\boldsymbol{q}\) is transformed to volume density through a density head \(\sigma_\theta\). Then, the network \(g_\theta(\cdot)\) operates on a shape encoding \(\boldsymbol{q}=(\gamma(\boldsymbol{x}),\boldsymbol{z}_s)\) that is later concatenated with the positional encoding of \(\boldsymbol{d}\) and appearance code \(\boldsymbol{z}_a\) (Fig. \ref{generator}c):
\begin{align}
    (\gamma(\boldsymbol{x}),\boldsymbol{z}_s) &\mapsto \boldsymbol{q}\\
    (\boldsymbol{q}(\boldsymbol{x},\boldsymbol{z}_s), \gamma(\boldsymbol{d}),\boldsymbol{z}_a) &\mapsto c\\
    \boldsymbol{q}(\boldsymbol{x},\boldsymbol{z}_s) &\mapsto \sigma
\end{align}

The final pixel response \(c_r\) is computed by the compositing operation (Fig. \ref{generator}c):
\begin{equation}
   c_r=\sum_{i=1}^Nc_r^i\alpha_r^i\exp{(-\sum_{j=1}^{i-1}\sigma_r^j\delta_r^j)} 
\end{equation}
where \(\alpha_r^i=1-\exp{(-\sigma_r^i\delta_r^i)}\) is the alpha compositing value of sampled point \(i\) and \(\delta_r^i=\parallel \boldsymbol{x}_r^{i+1}-\boldsymbol{x}_r^i\parallel_2\) is the distance between the adjacent sampled points.

In this way, both the density and pixel values are computed at each sampled point along the beam \(r\) with network \(g_\theta\). Finally, combining the results of all \(R\) beams, the generator \(G_\theta\) predicts an image patch \(\boldsymbol{P}_\textup{pred}\), as illustrated in Fig. \ref{generator}d.

\subsection{MedNeRF} \label{mednerf}
We investigate how we can adapt GRAF to the medical domain and apply it to render a volumetric representation from DRRs. Leveraging a large dataset, GRAF's discriminator \(D_\phi\) is able to continuously provide useful signals to train the generator \(G_\theta\). However, medical datasets like those considered in our problem are generally small, which causes two sequential issues:

\noindent\textbf{The lack of real information to the generator:} In GRAF (and in GAN in general), the only source of features of the training data contributing to the generator is the indirect gradient transferred from the discriminator. We find that the single convolutional feedback from GRAF’s discriminator poorly conveys refined features from DRRs resulting in inaccurate volumetric estimation.

\noindent\textbf{Brittle adversarial training:} With a limited training dataset, the generator or discriminator may fall into ill-posed settings such as mode collapse, which would lead to generating a limited number of instances and consequently, a suboptimal data distribution estimation. While some works have applied data augmentation techniques to leverage more data in the medical domain, some transformations could mislead the generator to learn the infrequent or even non-existent augmented data distribution \cite{tran_2020_data}. We find that naively applying classic data augmentation works less favorably than our adopted framework.\hfill\bigskip

\subsubsection{Self-supervised Learning for High-Fidelity Synthesis}
\label{subsubsec:self_supervised}

\begin{figure}[t]
\centering
\setlength\fboxsep{0pt}
\setlength\fboxrule{0.25pt}
\includegraphics[width=8.5cm]{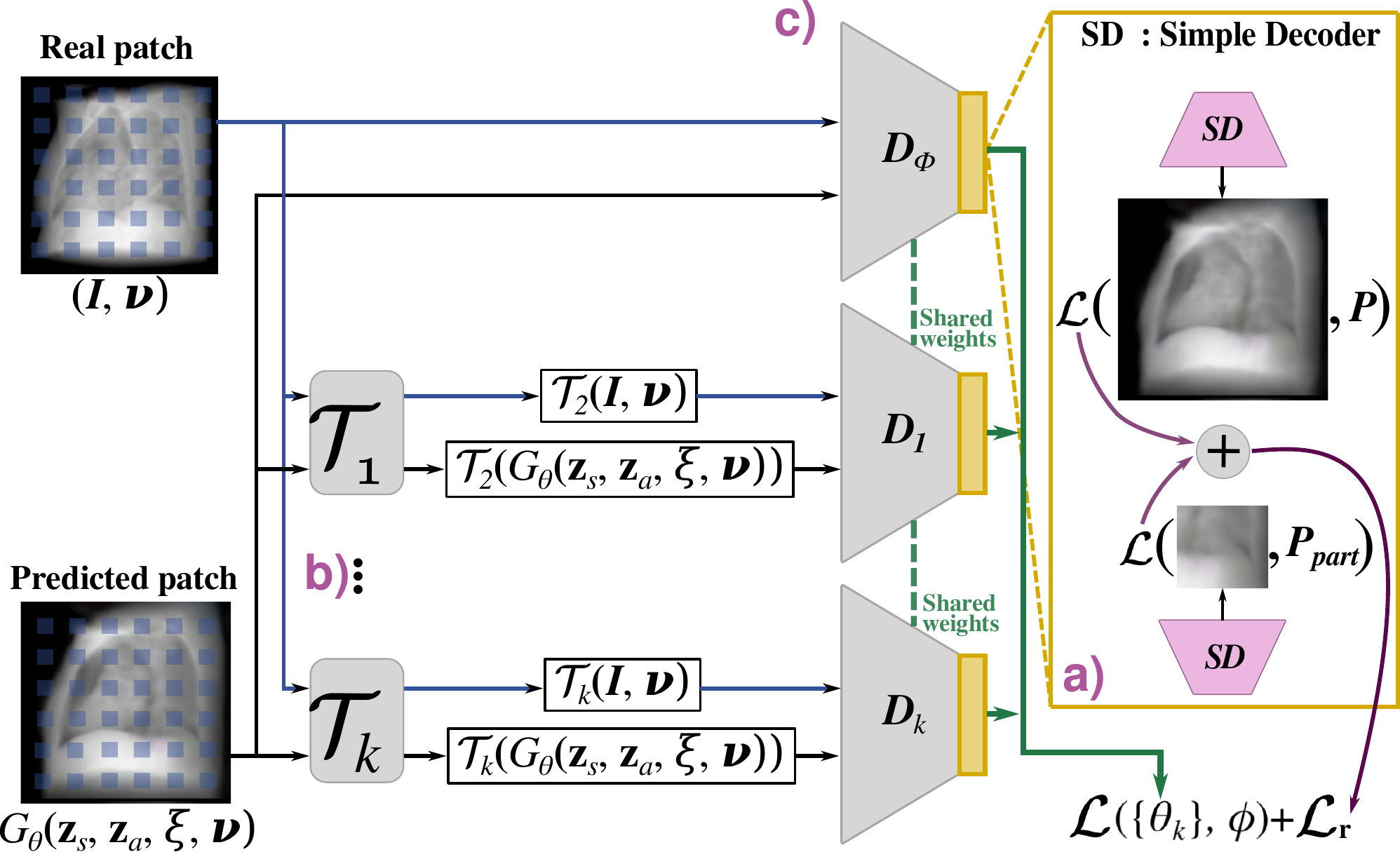}
\caption{An overview of our discriminator with self-supervised learning and DAG.} 
\label{discriminator}
\end{figure}

To allow \textit{richer} feature-maps covering from the DRRs such that it produces more comprehensive signals to train \(G_\theta\), we replace GRAF's discriminator architecture with recent advancements in self-supervised approaches. We allow \(D_\phi\) to learn useful global and local features training it on a pretext task, in particular, the self-supervision method based on auto-encoding \cite{liu_2021_towards}.  Different from \cite{liu_2021_towards}, we only use two decoders for the feature-maps on scales: \(\boldsymbol{f}_1\) on \(32^2\) and \(\boldsymbol{f}_2\) on \(8^2\) (Fig. \ref{discriminator}a). We find that this choice allows better performance and enables a correct volumetric depth estimation. \(D_\phi\) must therefore not only discriminate \(\boldsymbol{P}_\textup{pred}\) predicted from \(G_\theta\) but also extract comprehensive features from real image patches \(\boldsymbol{P}_\textup{real}\) that enable the decoders to resemble the data distribution. 

To assess global structure in decoded patches from \(D_\phi\), we use the Learned Perceptual Image Patch Similarity (LPIPS) metric \cite{Zhang_2018_lpips}. We compute the weighted pairwise image distance between two VGG16 feature spaces, where the pretrained weights are fit to better match human perceptual judgments. The additional discriminator loss is therefore:
\begin{equation}
    \mathcal{L}_\textup{r}=
     \mathbb{E}_{\boldsymbol{f}\sim\! D(\boldsymbol{p}), \, \boldsymbol{p}\sim\! \boldsymbol{P}}\left[\frac{1}{whd}\parallel\phi_i(\mathcal{G}(\boldsymbol{f}))-\phi_i(\mathcal{T}(\boldsymbol{p}))\parallel_2\right]
\end{equation}
where \(\phi_i(\cdot)\) denotes the \(i\)th layer output of a pretrained VGG16 network, and \(w\), \(h\), and \(d\) stand for the width, height and depth of a feature space, respectively.
Let \(\mathcal{G}\) be the processing on the intermediate feature-maps \(\boldsymbol{f}\) from \(D_\phi\), and \(\mathcal{T}\) the processing on real image patches. When coupled with this additional reconstruction loss, the network learns representations that transfer across tasks.\hfill\bigskip

\subsubsection{Improving Learning via Data Augmentation}\hfill
\label{subsubsec:data_aug}

We improve learning of \(G_\theta\) and \(D_\phi\) by adopting the Data Augmentation Optimized for GAN (DAG) framework \cite{tran_2020_data} in which a data augmentation transformation \(\mathcal{T}_k\) (Fig. \ref{discriminator}b) is applied using multiple discriminator heads \(\{D_k\}\). To further reduce memory usage, we share all layers of \(D_\phi\) except the last layers corresponding to each head (Fig. \ref{discriminator}c). Because applying differentiable and invertible data augmentation transformations \(\mathcal{T}_k\) has the Jenssen-Shannon (JS) preserving property \cite{tran_2020_data}:
\begin{equation}
    \textup{JS}(p_d^{\mathcal{T}_k}\parallel p_g^{\mathcal{T}_k}) = \textup{JS}(p_d\parallel p_g)
\end{equation}
where \(p_d^{\mathcal{T}_k}\) is the transformed training data distribution and \(p_g^{\mathcal{T}_k}\) the transformed distribution captured by \(G_\theta\). By using a total of four transformations combining flipping and rotation, we encourage optimization to the original data distribution, which also brings the most performance boost. These choices allow our model to benefit from not only \(\textup{JS}(p_d\parallel p_g)\) but also \(\textup{JS}(p_d^{\mathcal{T}_k}\parallel p_g^{\mathcal{T}_k})\), thereby improving the learning of \(G_\theta\) and generalization of \(D_\phi\). Furthermore, using multiple discriminators with weight-sharing provides learning regularization of \(D_\phi\).

Replacing GRAF's logistic objective with a hinge loss, we then define our overall loss as below:
\begin{equation}
    \mathcal{L}(\theta,\{\phi_k\})=\mathcal{L}(\theta,\phi_0)+\frac{\lambda}{n-1}\sum_{k=1}^n\mathcal{L}(\theta,\phi_k)
\end{equation}
\begin{equation}
\begin{split}
    \mathcal{L}&(\theta,\phi_k)= \\ &\mathbb{E}_{\boldsymbol{z}_s\sim p_s,\boldsymbol{z}_a\sim p_a, \boldsymbol{\xi}\sim p_\xi, \boldsymbol{\nu}\sim p_\nu}\left[f(D_\phi(G_\theta(\boldsymbol{z}_s,\boldsymbol{z}_a,\boldsymbol{\xi},\boldsymbol{\nu})))\right] \\
    &+\mathbb{E}_{\boldsymbol{I}\!\sim p_D, \boldsymbol{\nu}\sim p_\nu}\left[f(-D_\phi(\boldsymbol{I},\boldsymbol{\nu}))\right] + \mathcal{L}_\textup{r}
\end{split}
\end{equation}
where \(f(u)=\max(0,1+u)\). We optimize this loss with \(n=4\), where \(k=0\) corresponds to the identity transformation and \(\lambda=0.2\) (as in \cite{tran_2020_data}).\hfill\bigskip

\begin{figure*}[ht]
\centering
\setlength\fboxsep{0pt}
\setlength\fboxrule{0.25pt}
\includegraphics[width=7in]{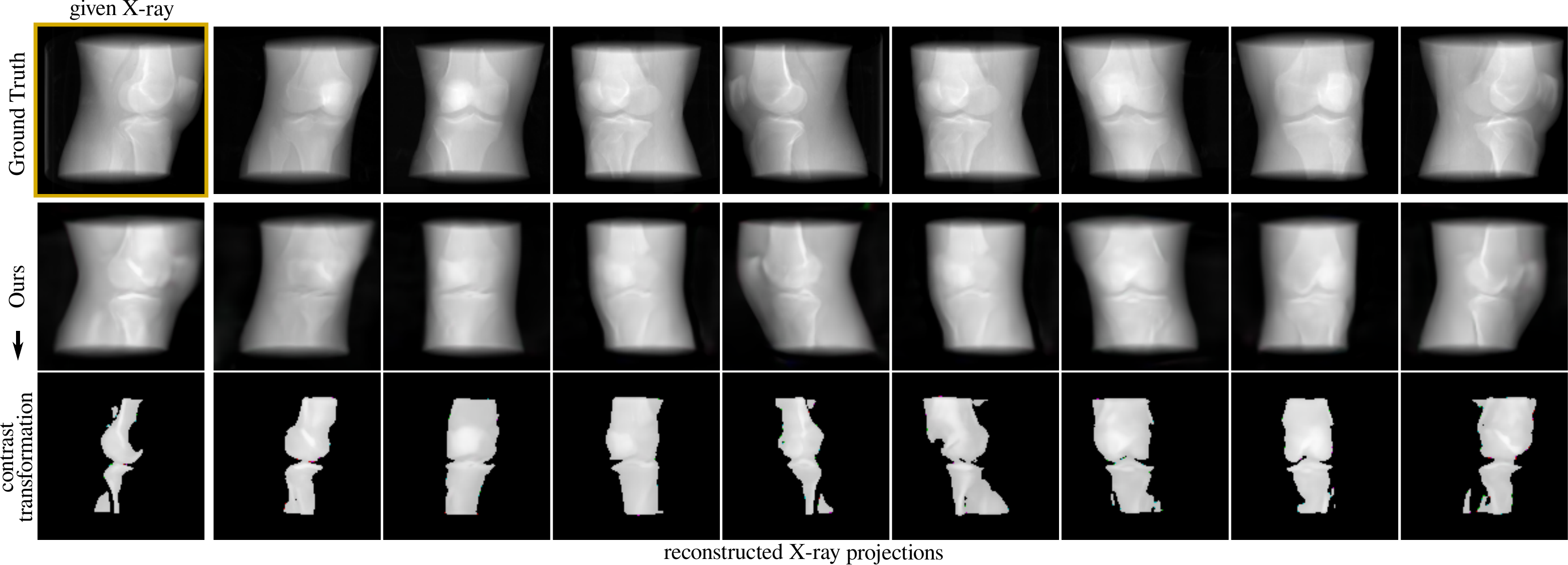}
\caption{Knee renderings from continuous viewpoint rotations showing tissue and bone. Given a single-view X-ray from a CT, we can generate the complete set of CT-projections within a full vertical rotation by \textit{slightly} fine-tuning a pretrained model along with the shape and appearance latent codes.}
\label{knee_recons}
\end{figure*}

\subsubsection{Volumetric Rendering from a Single View X-ray}\hfill
\label{subsubsec:single_recon}

After training a model, we reconstruct the complete X-ray projections within a full vertical rotation of a medical instance given a single view X-ray. We follow the relaxed reconstruction formulation in \cite{dgp_2021}, which fits the generator to a single image. Then, we allow the parameters of the generator \(G_\theta\) to be \textit{slightly} fine-tuned along with the shape and appearance latent vectors \(\boldsymbol{z}_s\) and \(\boldsymbol{z}_a\). The distortion and perception tradeoff is well known in GAN methods \cite{Blau_2018_CVPR} and therefore we modify our generation objective by adding the distortion Mean Square Error (MSE) loss, which incentivises a balance between blurriness and accuracy:
\begin{equation}
    \mathcal{L}_\textup{gen}=\lambda_\textup{1}\mathcal{L}_\textup{r}(VGG16)+\lambda_\textup{2}\mathcal{L}_\textup{MSE}(G)+\lambda_\textup{3}\mathcal{L}_\textup{NLLL}(\boldsymbol{z}_s,\boldsymbol{z}_a)
\end{equation}
where NLLL corresponds to the negative log-likelihood loss and the tuned hyperparameters \(lr=0.0005\), \(\beta_\textup{1}=0\), \(\beta_\textup{2}=0.999\), \(\lambda_\textup{1}=0.3\), \(\lambda_\textup{2}=0.1\) and \(\lambda_\textup{3}=0.3\).

Once the model locates an optimal combination of \(\boldsymbol{z}_s\) and \(\boldsymbol{z}_a\), we replicate them and use them to render the rest of the X-ray projections by continuously controlling the angle viewpoint.

\section{RESULTS}

\begingroup
\begin{table}
\caption {Quantitative results based on PSNR and SSIM of rendered X-ray projections with single-view X-ray input.}
\label{tab:recon_res}
\centering
\begin{tabular}{|r|c|c|}
\hline
 Dataset    & $\uparrow$ PSNR (dB) $(\mu\pm\sigma)$ & $\uparrow$ SSIM $(\mu\pm\sigma)$   
	\\    \hline  
Knee & $\mathbf{30.17}\pm1.93$ & $\mathbf{0.670}\pm0.040$ \\
\hline
Chest &  $\mathbf{28.54}\pm0.79$   & $\mathbf{0.462}\pm0.082$ \\ 
\hline
\end{tabular}
\end{table}
\endgroup

Here we provide an evaluation of MedNeRF on our datasets. We compare our model's results to the ground truth, two baselines, perform an ablation study, and show qualitative and quantitative evaluations. We train all models for 100,000 iterations with a batch size of 8. Projection parameters \((u, v)\) are chosen to evenly sample points on the surface of a sphere, specifically a slight horizontal elevation of 70-85 degrees and \(u_{\textup{min}} = 0\), \(u_{\textup{max}} = 1\) for a full 360-degree vertical rotation. However, we only provide a fifth of the views (72-views each at five degrees) during training and let the model render the rest.

\subsection{Reconstruction from Single View X-ray}
We evaluate our model's representation for 3D-aware DRR synthesis given a single-view X-ray as input. We find that despite the implicit linear network's limited capacity, our model can disentangle 3D anatomy identity and attenuation response of different medical instances, which are retrieved through the described reconstruction reformulation in \ref{subsubsec:single_recon}. Our model can also facilitate distinguishing bone from tissue via a contrast transformation, as it renders a brighter pixel value for denser structures (e.g. bone) (Fig. \ref{knee_recons}).

Table \ref{tab:recon_res} summarises our results based on the peak signal-to-noise ratio (PSNR) and structural similarity (SSIM), which measure the quality of reconstructed signals and human subjective similarity, respectively. We find that our generative loss can achieve a reasonable perception-distortion curve in renderings and show consistency with the location and volumetric depth of anatomical structures at continuous viewpoints compared to the ground truth.

\begin{figure}[t]
\centering
\setlength\fboxsep{0pt}
\setlength\fboxrule{0.25pt}
\includegraphics[width=8.5cm]{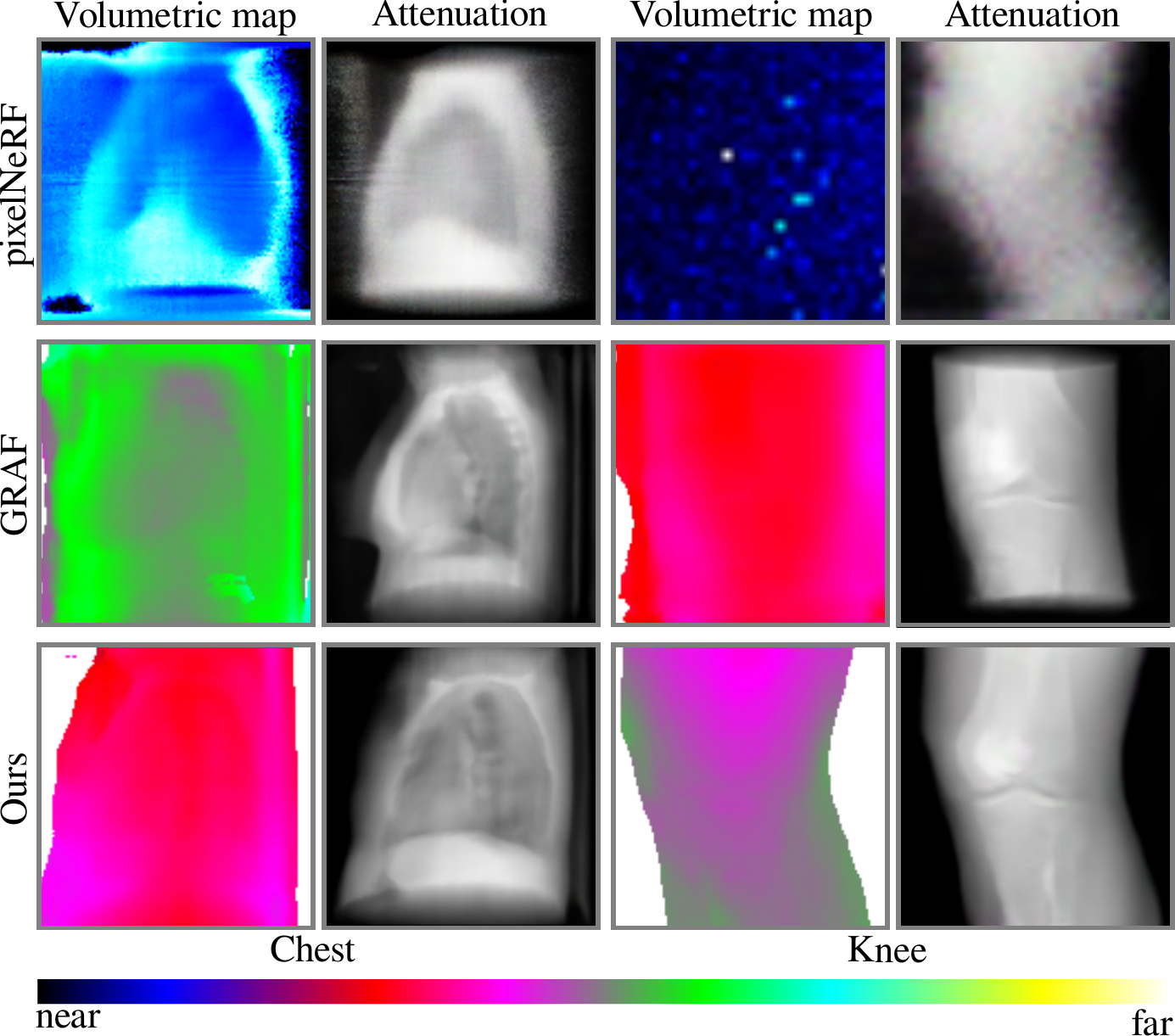}
\caption{Volumetric maps and attenuation renderings on our dataset.}
\label{qual_res}
\end{figure}

\subsection{2D DRR Rendering}
We evaluate our model on the task of 2D rendering and compare it to pixelNeRF \cite{yu2020pixelnerf}, and GRAF \cite{Schwarz2020NEURIPS} baseline, wherein the original architecture is used. Our model can more accurately estimate volumetric depth compared to GRAF and pixelNeRF (Fig. \ref{qual_res}). For each category, we find an unseen target instance with a similar view direction and shape. Volumetric depth estimation is given by bright colors (far) and dark colors (near). Lacking a perceptual loss, GRAF is not incentivized to produce high-frequency textures. In contrast, we find our model renders a more detailed internal structure with varied attenuation. GRAF produces a consistent attenuation response, but seems to be unable to distinguish the anatomical shape from the background. Our self-supervised discriminator enables the generator to disentangle shape and background by rendering a brighter color for the background and a darker color for the shape, while GRAF renders a bright or dark color for both.

We find pixelNeRF produces blurred attenuation renderings for all datasets, and volumetric maps tend to exhibit strong color shifts (Fig. \ref{qual_res}). We believe these artifacts are due to the see-through nature of the dataset, compared to solid-like natural objects on which NeRFs are trained. This data characteristic impairs not only volumetric maps but also fine anatomical structures. In contrast, our model is better able to render both volumetric depth and attenuation response. We also find pixelNeRF is sensitive to slight changes in projection parameters, hampering optimization for the knee category. Our model produces a consistent 3D geometry and does not rely on explicit projection matrices. 

\begingroup
\tiny
\begin{table}
\caption {FID and KID analysis comparing other methods.}
\label{tab:quant_res}
\resizebox{\columnwidth}{!}{%
\centering
\begin{tabular}{|r|c|c|c|c|}
\hline
 & \multicolumn{2}{c|}{Chest dataset} & \multicolumn{2}{c|}{Knee dataset}  \\
\cline{2-3} \cline{4-5}
 Method    & $\downarrow$ FID $(\mu\pm\sigma)$ & $\downarrow$ KID $(\mu\pm\sigma)$   
	& $\downarrow$ FID $(\mu\pm\sigma)$ & $\downarrow$ KID $(\mu\pm\sigma)$  
	\\    \hline  
GRAF \cite{Schwarz2020NEURIPS} & $68.25\pm0.954$ & $0.053\pm0.0008$ & $76.70\pm0.302$ & $0.058\pm0.0001$ \\
pixelNeRF \cite{yu2020pixelnerf} &  $112.96\pm2.356$   & $0.084\pm0.0012$ &  $166.40\pm2.153$   & $0.158\pm0.0010$    \\
Ours &  $\mathbf{60.26\pm0.322}$   & $\mathbf{0.041\pm0.0005}$ &  $\mathbf{76.12\pm0.193}$   & $\mathbf{0.052\pm0.0004}$  \\   
\hline
\end{tabular}}
\end{table}
\endgroup

\begingroup
\tiny
\begin{table}
\caption {FID and KID analysis of ablations of our model.}
\label{tab:ablations}
\resizebox{\columnwidth}{!}{%
\centering
\begin{tabular}{|r|c|c|}
\hline
 & \multicolumn{2}{c|}{Chest dataset}  \\
\cline{2-3}
Ablation	& $\downarrow$ FID $(\mu\pm\sigma)$ & $\downarrow$ KID $(\mu\pm\sigma)$  
	\\    \hline 
MedNeRF - 3 SD, logistic loss, classic DA &  $84.85\pm1.025$   & $0.069\pm0.0031$ \\
MedNeRF - 2 SD, logistic loss, classic DA &  $67.73\pm0.712$   & $0.051\pm0.0006$ \\
MedNeRF - 2 SD, hinge loss, classic DA & $65.34\pm0.353$   & $0.045\pm0.0004$  \\ 
MedNeRF - 2 SD, hinge loss, DAG & $\mathbf{60.26\pm0.322}$   & $\mathbf{0.041\pm0.0005}$ \\
\hline
\end{tabular}}
\end{table}
\endgroup

Table \ref{tab:quant_res} compares image quality based on Frechet Inception Distance (FID) and Kernel Inception Distance (KID) metrics, in which lower values mean better. Optimizing pixelNeRF on our datasets leads to particularly poor results that are unable to compete with the GRAF baseline and our model. In contrast, our model outperforms the baselines on FID and KID metrics for all datasets. 

\subsection{Ablation Study}
We evaluate our model with three ablations (Table \ref{tab:ablations}): wherein an additional simple decoder (SD) is included; the adversarial logistic loss is replaced by its hinge version; and wherein the non-classical DAG approach is adopted. We find that that the DAG approach brings the most performance boost compared to naively applying classical DA, while the use of a hinge loss performs slightly better than its logistic version. However, an additional decoder in our self-supervised discriminator can lead to a significant drop in performance.

\section{CONCLUSION}

We have presented a novel Deep Learning architecture based on Neural Radiance Fields for learning a continuous representation of CT scans. We learn a medical category encoding of the attenuation response of a set of 2D DRRs in the weights of a generator. Furthermore, we have found that a stronger and more comprehensive signal from our discriminator allows generative radiance fields to model 3D-aware CT-projections. Experimental evaluation demonstrates significant qualitative and quantitative reconstructions and improvements over other Neural Radiance Field approaches. Whilst the proposed model may not replace CT entirely, the functionality of generating 3D-aware CT-projections from X-rays has great potential for clinical use in osseous trauma, skeletal evaluation in dysplasia and for orthopaedic pre-surgical planning. This could cut down on the radiation dose given to patients, with significant economic implications such as bringing down the cost of investigations.

\addtolength{\textheight}{-12cm}   




\section*{ACKNOWLEDGMENT}
This work is partially supported by the Mexican Council of
Science and Technology (CONACyT).


\bibliographystyle{IEEEbib}
\bibliography{refs}

\end{document}